# Ventricle features as reliable differentiators between the bvFTD and other dementias

Ana L. Manera[1], M.D., Mahsa Dadar[1,2], Ph.D, D. Louis Collins[1], Ph.D.*, Simon Ducharme[1,3], M.D. MSc*

Frontotemporal Lobar Degeneration Neuroimaging Initiative (FTLDNI)†

Alzheimer's Disease Neuroimaging Initiative‡

[1] McConnell Brain Imaging Centre, Montreal Neurological Institute, McGill University, Montreal, Quebec (QC), Canada.

[2] CERVO Brain Research Center, Centre intégré universitaire santé et services sociaux de la Capitale Nationale, Québec, QC

3 Department of Psychiatry, Douglas Mental Health University Health Centre, McGill University, Montreal, Quebec (QC), Canada.

*This is a shared senior authorship

† Data used in preparation of this article were obtained from the Frontotemporal Lobar Degeneration Neuroimaging Initiative (FTLDNI) database (http://4rtni-ftldni.ini.usc.edu/ ). The investigators at NIFD/FTLDNI contributed to the design and implementation of FTLDNI and/or provided data but did not participate in analysis or writing of this report (unless otherwise listed).

‡Part of the data used in preparation of this article was obtained from the Alzheimer's Disease Neuroimaging Initiative (ADNI) database (adni.loni.usc.edu). As such, the investigators within the ADNI contributed to the design and implementation of ADNI and/or provided data but did not participate in analysis or writing of this report. A complete listing of ADNI investigators can be found at:

http://adni.loni.usc.edu/wpontent/uploads/how_to_apply/ADNI_Acknowledgement_List.pdf.

**Corresponding Author Information:**

Ana L. Manera, Montreal Neurological Institute, 3801 University Street, Room WB320, Montréal, QC, H3A 2B4

Email: ana.manera@mail.mcgill.ca

**Abstract**

INTRODUCTION: Lateral ventricles are reliable and sensitive indicators of brain atrophy and disease progression in behavioral variant frontotemporal dementia (bvFTD). We aimed to investigate whether an automated tool using ventricular features could improve diagnostic accuracy in bvFTD across neurodegenerative diseases.

METHODS: Using 678 subjects (2750 timepoints), differences in ventricular features were assessed between bvFTD, normal controls and other dementia cohorts.

RESULTS: Ventricular antero-posterior ratio (APR) was the only feature that was significantly different and increased faster in bvFTD compared to all other cohorts. We achieved a 10-fold cross-validation accuracy of 80% (77% sensitivity, 82% specificity) in differentiating bvFTD from all other cohorts with ventricular features, and 76% accuracy using only the single APR feature.

DISCUSSION: APR could be a useful and easy-to-implement feature to aid bvFTD diagnosis. We have made our ventricle feature estimation and bvFTD diagnostic tool publicly available, allowing application of our model in other studies.

**Abbreviations**

bvFTD= behavioral variant frontotemporal dementia

MRI= magnetic resonance imaging

DBM= deformation-based morphometry

CN= cognitively normal controls

AD= Alzheimer's disease

MCI= mild cognitive impairment

FTD= frontotemporal dementia

SV= semantic variant

PNFA= progressive non fluent aphasia

ADNI=Alzheimer's Disease Neuroimaging Initiative

NIFD= Frontotemporal Lobar Degeneration Neuroimaging Initiative

T1w= T1-weighted

FDR= False Discovery Rate

APR= anteroposterior ratio

AUC= area under the curve

**Introduction**

In the absence of a pathologic genetic mutation, the diagnostic certainty of the behavioral variant frontotemporal dementia (bvFTD) still relies on the convergence of clinical criteria and structural magnetic resonance imaging (MRI) or nuclear medicine imaging findings. In recent work, using deformation-based morphometry (DBM), we showed that the ventricles play a remarkable role in discriminating bvFTD from cognitively normal controls (CN) and their expansion proved to be a sensitive indicator of disease progression [1,2]. Although ventricular enlargement is not specific to bvFTD, it has been found that it has higher rates of expansion than Alzheimer's disease (AD) [3-5].

Given the sensitivity limitations of brain MRI in the early stages of FTD, there is an increasing interest in the development of automated quantitative and volumetric tools for MRI to improve diagnostic accuracy[6,7]. Most of these efforts have focused on cortical atrophy measurements, however this is hard to reliably perform at the individual subject level in clinical settings, and tools based on specific cortical regions do not account for inter-subject variability in atrophy distributions. As opposed to the cortical surface, the lateral ventricles are easy to reliably segment manually or when using standard publicly available tools [8,9] and provide an estimate of the overall extent of brain atrophy across different regions, making ventricle-based features for bvFTD diagnosis a promising practical tool both in research and clinical settings. The aim of this study was to further investigate the relevance of assessing ventricle enlargement and shape features in differentiating bvFTD from other dementias, findings which are readily available from MRIs in current practice but that insufficiently used. We performed surface and volumetric analysis on lateral ventricles in order to find a reliable differentiator of bvFTD from AD, mild cognitive impairment (MCI), the language variants of frontotemporal dementia -FTD- (Semantic Variant -

SV- and Progressive Nonfluent Aphasia -PNFA) and CN, with the goal of developing a clinically usable tool.

**Materials and Methods**

**Participants**

This study included 678 subjects - 69 bvFTD, 38 SV, 37 PNFA, 218 MCI, 74 AD and 242 CN with a total of 2750 timepoints from the Alzheimer's Disease Neuroimaging Initiative (ADNI) and the Frontotemporal Lobar Degeneration Neuroimaging Initiative (FTLDNI).

The FTLDNI was funded through the National Institute of Aging and started in 2010. The primary goals of FTLDNI are to identify neuroimaging modalities and methods of analysis for tracking frontotemporal lobar degeneration and to assess the value of imaging versus other biomarkers in diagnostic roles. The project is the result of collaborative efforts at three sites in North America. For up-to-date information on participation and protocol, please visit: http://4rtni-ftldni.ini.usc.edu/. Data was accessed and downloaded through the LONI platform in August 2018. We included all 69 bvFTD, 38 SV and 37 PNFA patients and 129 age matched CNs from the FTLDNI database who had T1-weighted (T1w) MRI scans matching with each clinical visit (Table 1).

The ADNI dataset was launched in 2003 as a public–private partnership, led by Principal Investigator Michael W. Weiner, MD. The primary goal of ADNI has been to test whether serial MRI, positron emission tomography, other biological markers, and clinical and neuropsychological assessment can be combined to measure the progression of MCI and AD.

ADNI was carried out with the goal of recruiting 800 adults aged from 55 to 90, and consists of 200 cognitively normal, 400 MCI, and 200 AD subjects. ADNIGO is a later study that followed ADNI participants that were in cognitively normal or early MCI stages (http://www.adcs.org/studies/imagineadni.aspx). ADNI2 study followed patients in the same categories as well as recruiting 550 new subjects (http://www. adcs.org/studies/ImagineADNI2.aspx).

In the present study, we included 74 AD amyloid β+ (defined based on the composite scores from UC Berkeley AV45 assessments provided by the ADNI with a normalized cut off threshold of 0.79), 218 MCI amyloid β+ and 113 CNs from ADNI with T1w MRI scans matching with each clinical visit and age matched to the FTLDNI cohort (Table 1). The 74 AD, 218 MCI and 113 CN represent all subjects from ADNI where this data was available, randomly selected to age match the FTLDNI cohort. All subjects included provided informed consent and the protocol was approved by the institution review board at all sites.

**Neuroimaging**

**Image acquisition and preprocessing**

For the FTLDNI cohort, 3.0T MRIs were acquired at three sites (T1w MPRAGE, TR=2 ms, TE=3 ms, IT=900 ms, flip angle 9°, matrix 256x240, slice thickness 1mm, voxel size 1mm$^3$). Within ADNI, T1w scans from ADNI1 dataset were acquired in 3D with a gradient recalled sequence with 1.2 mm slice thickness, 160 sagittal slices, a 192 × 192 mm field of view, and a 192 × 192 scan matrix, voxel size of 1.2 × 0.9375 ×0.9375 mm, TR = 3000 ms, and TE = 3.55 ms. For ADNI2/GO datasets, T1w scans were acquired in 3D with a gradient recalled sequence with 1.2 mm slice thickness, 196 sagittal slices, covering the entire brain, a 256 × 256 mm field of view, and a 256 × 256 scan matrix, voxel size of 1 × 1 × 1.2 mm, TR = 7.2 ms, and TE = 3.0 ms.

The T1w scans of the subjects were pre-processed through our longitudinal pipeline [10] that included image denoising [11], intensity non-uniformity correction [12], and image intensity normalization into range (0−100) using histogram matching. Each native T1w volume from each timepoint was linearly registered first to a subject-specific template which was then registered to the ICBM152-2009c template [13]. The images were visually assessed by two experienced raters to exclude cases with significant imaging artifacts (e.g. motion, incomplete field of view) or inaccurate linear/nonlinear registrations. This visual assessment was performed blind to diagnosis.

**Ventricle segmentation**

A previously validated patch-based label fusion technique was employed to segment the lateral ventricles [9]. The method uses expert manual segmentations as priors and estimates the label of each test subject voxel by comparing its surrounding patch against all the patches from the training library and performing a weighted label fusion using the intensity-based distances between the patch under study and the patches in the training subjects. All resulting segmentations were visually assessed and the incomplete/inaccurate segmentations (N=23 subjects/30 scans) were manually corrected by an experienced rater. The process of segmentation QC and manual correction was performed blind to the clinical diagnosis.

**Ventricular volume and shape features estimation**

Using a lobe atlas of the brain delineating frontal, parietal, temporal, and occipital lobes separately in the left and right hemispheres based on the Hammers atlas [14, 15], lateral ventricular volumes were calculated per each brain lobe and hemisphere. Using coronal coordinate y=12mm in the stereotaxic space (i.e. registered to the template) ventricles were divided into anterior and posterior portions. All volumes were normalized for intracranial volume and these ratios were log-transformed to achieve normal distribution. Surface and surface to volume ratio estimations where also obtained.

**Deformation based morphometry**

DBM was used to assess voxel-wise group related volumetric differences in the ventricles. Each individual scan was nonlinearly registered to the ICBM152-nonlin_sym_2009c [16] template using the ANTs diffeomorphic registration pipeline [17]. The integral of the Jacobian determinant of the inverse deformation field from the non-linear transformations within the lateral ventricle was used as a measure of ventricle expansion or shrinkage. Local contractions can be interpreted as shrinkage and local expansions as enlargement of the region.

**Statistical Analysis**

All statistical analyses were conducted using MATLAB (version R2019b). Differences in categorical variables between the cohorts were assessed using chi-square and continuous variables were assessed using one-way ANOVA or Kruskal-Wallis variance analysis depending on the distribution of the variables based on normality test. Post-hoc two-sample t-Tests were conducted to examine clinical and imaging differences at baseline. Results are expressed as mean ± standard deviation and median [interquartile range] as appropriate.

**Voxel-wise Analysis:** Voxel-wise mixed effects model analysis was performed to determine the patterns of differences between each cohort and their age matched controls:

$DBM \sim 1 + Cohort + Age + Sex + 1|ID$

where DBM denotes voxel-wise Jacobian values for subject timepoints. The variable of interest was Cohort, a categorical fixed variable contrasting each cohort versus CN. The resulting maps were corrected for multiple comparisons using False Discovery Rate (FDR) controlling method, with a significance threshold of 0.05.

**Feature Analyses:** Similarly, longitudinal mixed-effects models were used to assess the slope differences (with regards to changes with age) in ventricular features (anteroposterior ventricular ratio -APR- and total ventricular volume -TVV-) between bvFTD and age-matched controls as well as other dementia cohorts.

$Model\ feature \sim 1 + Cohort + Age + Cohort:Age + Sex + 1|ID$

The variable of interest was the interaction between Cohort and Age, denoted by Cohort:Age. In all mixed-effects models, subject ID was considered as categorical random effects. The models also included sex as a categorical fixed variable.

**Diagnosis Classification:** To further demonstrate the diagnostic relevance of the ventricle-based features in differentiating between bvFTD and other cohorts, the features were used alone and also in combination with each other, age, and sex, to differentiate bvFTD from all other cohorts. A support vector machine classifier was trained on each feature set (fitcsvm function from MATLAB, with default parameters: linear kernel, Sequential Minimal Optimization) to perform the classification task, balancing the number of subjects included in each class. A 10-fold cross validation scheme was employed, and the process was repeated 100 times to obtain a robust estimate of the performance of the classifier.

**RESULTS**

**Demographics**

Table 1 shows the demographic and cognitive testing performances for all the cohorts. bvFTD, SV and CN$_{NIFD}$ subjects were younger than PNFA, CN$_{ADNI}$ and amyloid+ MCI and AD. The median follow up time in years was significantly longer for CN and MCI than AD and all the FTD related cohorts. In general, for all the cognitive/functional scores assessed CN subjects performed, as expected, significantly better than the all the other groups. There were no significant differences between bvFTD and AD cohorts in MMSE, MoCA and CDR-SB and they both performed significantly worse than the rest of the cohorts in all the measures. Finally, MCI subjects did not show significant differences in MMSE and MoCA scores with SV and PNFA cohorts and they had lower CDR-SB scores than AD, bvFTD and SV subjects.

|  | NIFD | | | | ADNI | | | P |
| --- | --- | --- | --- | --- | --- | --- | --- | --- |
|  | CN | bvFTD | SV | PNFA | CN | MCI | AD | |
|  | (N=129) | (N=69) | (N=38) | (N=37) | (N=113) | (N=218) | (N= 74) | |
| Scans | 447 | 231 | 175 | 138 | 500 | 1037 | 222 | |
| Age (y) | 62±7 | 61±6 | 62±6 | 68±7 | 70±4 | 69±5 | 69±6 | <0.001 |
| Sex (male%) | 56 (43%) | 45 (65%) | 21 (55%) | 17 (46%) | 64 (57%) | 113 (52%) | 34 (46%) | 0.07 |
| Follow-up (y) | 1.3[0.6-3.4] | 1[0.5-1.4] | 1.2[1-16]] | 1.1[0.6-1.5] | 2[1.1-2.3] | 2[1.1-3] | 1[0.2-1.1] | <0.001 |
| MMSE | 30[29-30] | 25[22-27] | 26[22-28] | 27[21-28] | 29[29-30] | 28[27-29] | 24[21-25] | <0.001 |
| MoCA | 28[25-29] | 19[12-23] | 21[17-22] | 21[10-25] | 26[25-28] | 23[21-26] | 19[14-21] | <0.001 |
| CDR-SB | 0 | 6[4.5-8.5] | 3.5[2.5-5.5] | 2[1-3.5] | 0 | 1.5[1-2] | 4.5[3-5] | <0.001 |

**Table 2.** Demographic and clinical characteristics for all the cohorts. Values express Mean± SD / Median [interquartile range]. P value level of significance: 0.05. Abbreviations: CN: cognitively normal controls; bvFTD: behavioral variant frontotemporal dementia; SV: semantic variant; PNFA: primary nonfluent aphasia; MCI: mild cognitive impairment; AD: Alzheimer's dementia.

**Baseline ventricular volumes**

Figure 1 shows the statistically significant local differences in ventricular volumes between each cohort and CN after FDR correction, plotted on top of the average ADNI template[18]. Warmer colors indicate greater differences (i.e. greater degrees of ventricular enlargement). While some degree of ventricular enlargement was found for all cohorts, bvFTD showed the greatest difference in ventricular volume compared to CN; as seen in Fig. 1 the dark red anterior horns of the lateral ventricle in the bvFTD are roughly 10 time larger than the controls.

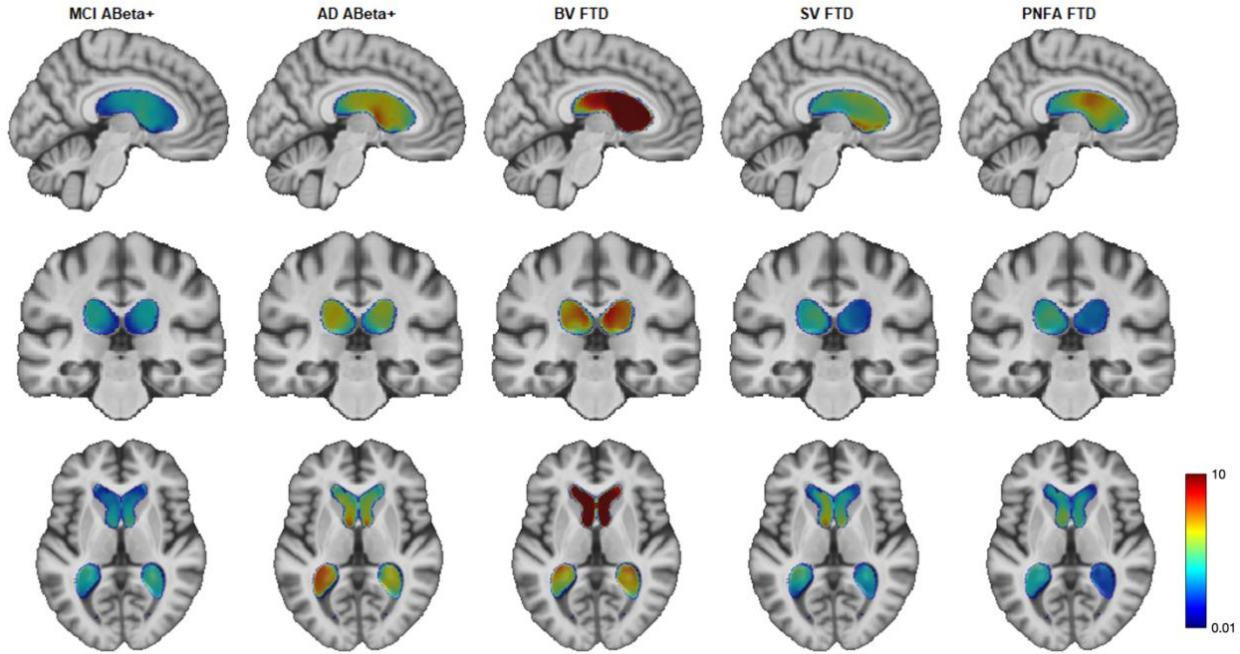

**Figure 1.** Voxel-wise DBM Jacobian beta maps indicating significant ventricular differences between each cohort and age-matched controls (FDR corrected p-value<0.05). From left to right: MCI vs Controls, AD vs Controls, bvFTD vs Controls, PNFA vs Controls and SV vs Controls. Model: $DBM_{Ventricles} \sim 1 + Dx + AGE + Dx:AGE + Sex + (1|ID) + (1|SITE)$. The figures show the significant beta values obtained for the categorical variable DX (i.e. bvFTD vs Controls). Colormaps within the ventricles shows the degree of ventricle enlargement for each cohort compared to controls overlaid on the ADNI unbiased average brain template (warmer colors indicate regions with greater ventricle enlargement than regions with colder colors, varying from 1% larger to 10 times larger).

The differences in lobar and total ventricular volumes between the cohorts are shown in Figure 2. These results demonstrate an important overlap in VV between cohorts. Figure 3 shows the left/right hemisphere ratios per lobe for the different cohorts (Panel 3A) and the APR comparison between cohorts (Panel 3B). The ventricular APR was significantly larger for bvFTD compared to all other cohorts ($APR_{bvFTD}$ 1.4±0.5, $APR_{Control}$ 1±0.2, $APR_{MCI}$ 0.97±0.2, $APR_{AD}$ 0.92±0.22, $APR_{SV}$ 1.1±0.3, $APR_{PNFA}$ 1.2±0.5; p<0.001). Mean volumes and ratios for all the cohorts are shown in Table 2.

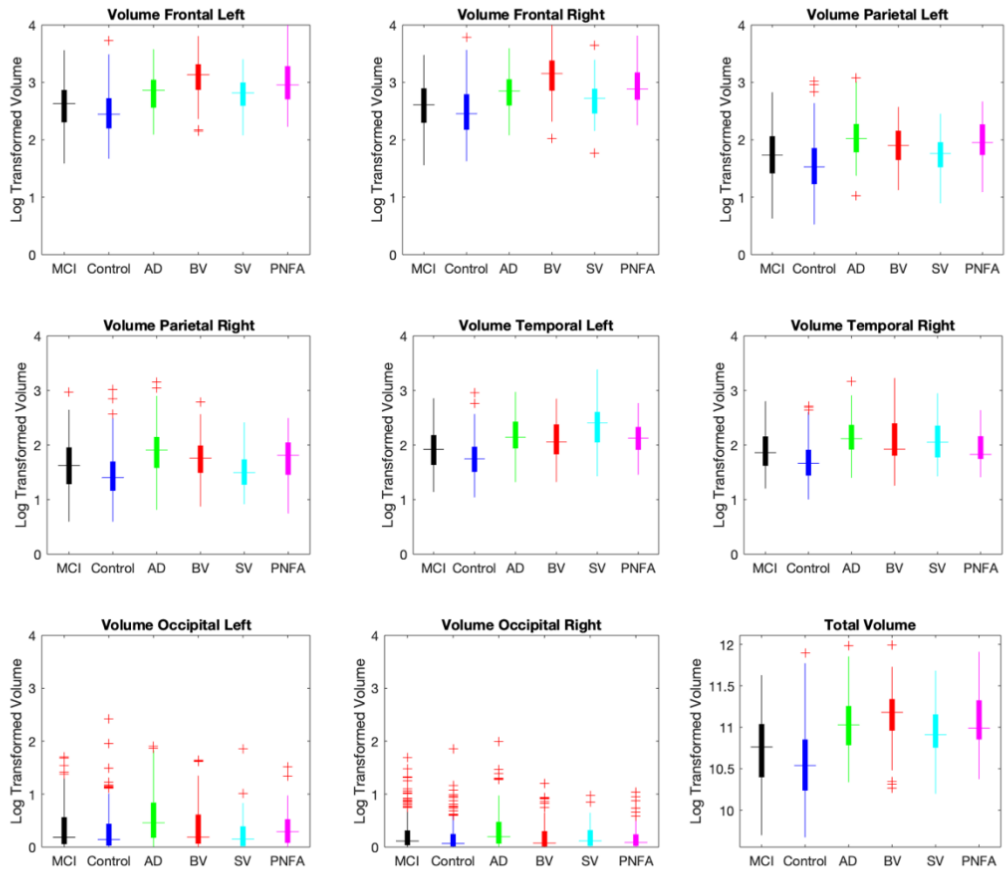

**Figure 2**. Log-transformed Ventricular volume per lobe, and total venricular volume, for different cohorts. Volumes (ml) shown are log-transformed, note that a log transformed volume of 3 corresponds to a volume of 1000 cc before log transformation.

Abbreviations: MCI: mild cognitive impairment; AD: Alzheimer's disease; BV: behavioral variant; SV: semantic variant; PNFA; progressive non-fluent aphasia.

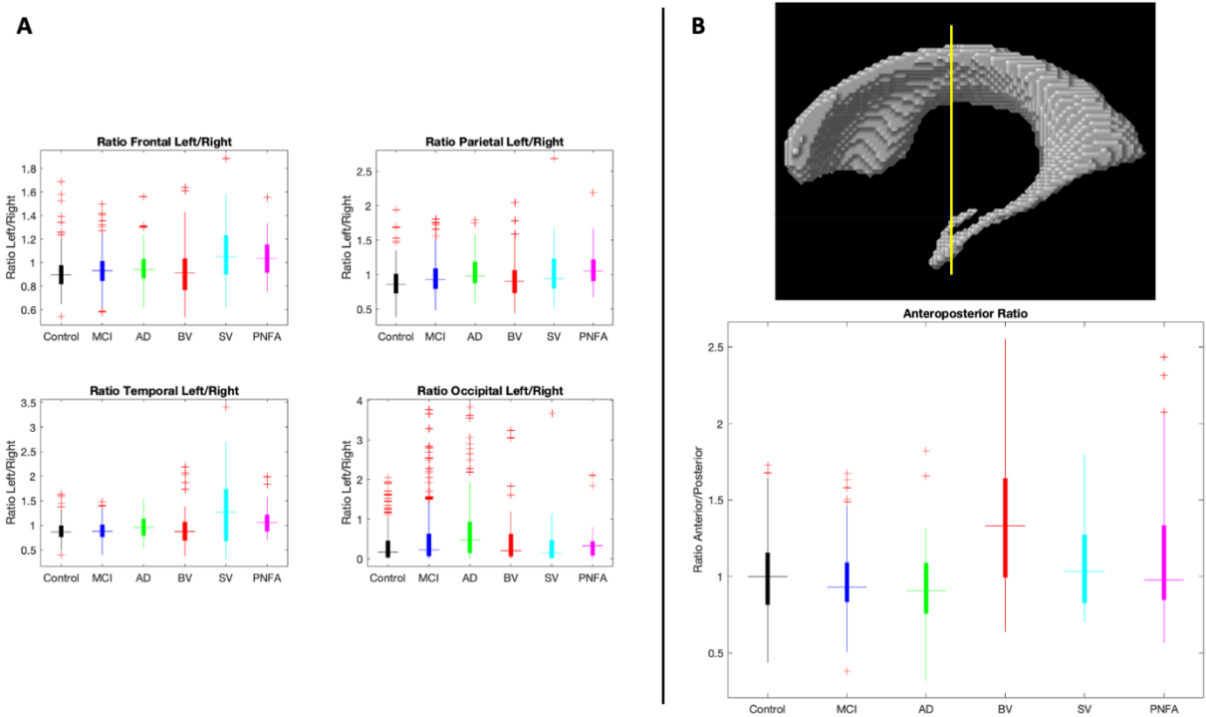

**Figure 3. Panel A:** Volume left/right ratios per lobe comparison between cohorts. **Panel B:** Upper figure: coronal coordinate y=12mm for anteroposterior ratio estimation. Lower figure: Anteroposterior ratio for different cohorts (anterior ventricular volume / posterior ventricular volume)

Abbreviations: MCI: mild cognitive impairment; AD: Alzheimer's disease; BV: behavioral variant; SV: semantic variant; PNFA; progressive non-fluent aphasia.

| Feature | Controls | MCI | AD | BV | SV | PNFA | P | Group Differences |
|---|---|---|---|---|---|---|---|---|
| Right frontal volume | 12.1±5.8 | 13.5±5.8 | 17±6.4 | 23.5±9.7 | 15.2±6.4 | 19.2±8.1 | <0.001 | **BV**; PNFA > AD; SV; MCI; Control |
| Left Frontal volume | 11.9±5.5 | 13.4±5.6 | 17±6.4 | 22.2±8 | 16.5±5.4 | 21.2±9.7 | <0.001 | **BV**; PNFA > AD; SV; MCI; Control |
| Right Parietal Volume | 3.8+2.6 | 4.5±2.7 | 6.3±3.6 | 5.4±2.7 | 3.9±2 | 5.2±2.5 | <0.001 | AD; **BV**; PNFA > MCI; SV; Control |
| Left Parietal volume | 4.3±2.7 | 5.3±2.9 | 7.5±3.8 | 5.9±2.5 | 4.9±2.1 | 6.7±2.9 | <0.001 | AD; PNFA > **BV**; SV; MCI; Control |
| Right temporal volume | 4.8±.2.1 | 6.1±2.7 | 8±3.5 | 7.4±3.7 | 7.8±4 | 6.3±2.5 | <0.001 | AD; SV ;**BV**> PNFA; MCI > Conrtrol |
| Left temporal volume | 5.1±2.2 | 6.2±2.7 | 8.5±3.5 | 7.6±3.2 | 10.3±4.6 | 7.8±2.9 | <0.001 | SV; AD ;PNFA> **BV**; MCI > Control |
| Right occipital volume | 0.2±0.5 | 0.4±0.6 | 0.6±1 | 0.6±0.9 | 0.5±0.9 | 0.5±0.8 | <0.001 | AD;MCI> **BV**; SV; PNFA; Control |
| Left occipital volume | 0.4±0.9 | 0.6±0.8 | 1.1±1.4 | 0.68±1.05 | 0.61±1.06 | 0.57±0.82 | <0.001 | AD;PNFA> **BV**; SV; MCI; Control |
| Anterior volume | 20.9±9.6 | 24.1±9.8 | 30.7±11.3 | 41.1±16.2 | 30.1±10.6 | 35.2±16.4 | <0.001 | **BV**; PNFA; AD > SV; MCI > Control |
| Posterior volume | 21.8±11.4 | 25.9±12.6 | 35.4±16.4 | 31.8±12.2 | 29/4±10.9 | 32±11.8 | <0.001 | AD; **BV**; SV; PNFA > MCI; Control |
| Total volume | 42726+20243 | 49992±21475 | 66101±26289 | 72905±25940 | 59444±20277 | 67246±24955 | <0.001 | **BV**; AD; SV; PNFA > MCI, Control |
| Left/right frontal ratio | 1+0.2 | 1+0.2 | 1+0.2 | 1+0.3 | 1.2±0.3 | 1.1±0.2 | <0.001 | SV; PNFA > AD; **BV**; MCI; Control |
| Left/right Parietal ratio | 1.2±0.3 | 1.2±0.3 | 1.3±0.3 | 1.2±0.4 | 1.4±0.5 | 1.2±0.4 | 0.001 | PNFA; SV ; AD; MCI; **BV** > Control |
| Left/right Temporal ratio | 1.1±0.2 | 1.1±0.2 | 1.1±0.3 | 1.1±0.5 | 1.5±0.8 | 1.3±0.4 | <0.001 | SV; PNFA> AD; **BV**; MCI; Control |
| Anteroposterior ratio | 1±0.2 | 1±0.2 | 0.9±0.2 | 1.4±0.5 | 1.1±0.3 | 1.2±0.5 | <0.001 | **BV**> AD; SV; PNFA; MCI; Control |

**Table 2.** Lobar ventricular volumes and ratios for all the cohorts. Values express Mean± SD. P value level of significance: 0.05. Group differences are shown in the last column.

**Longitudinal anteroposterior ratio change**

Figure 4 shows the slope differences (changes with age) in ventricular APR and total VV between bvFTD and age-matched controls as well as other dementia cohorts. Dotted lines indicate confidence intervals of the estimated lines. While total ventricle volume becomes larger with age in all cohorts, bvFTD patients show a much faster increase in the ventricular antero-posterior ratio compared to all other cohorts ($p \leq 0.01$). This trend of increase in the ventricular antero-posterior ratio is specific to the bvFTD group; in contrast, the antero-posterior ratio decreases in control, MCI, and AD subjects, remains relatively stable for SV subjects and increases at a slower pace in PNFA.

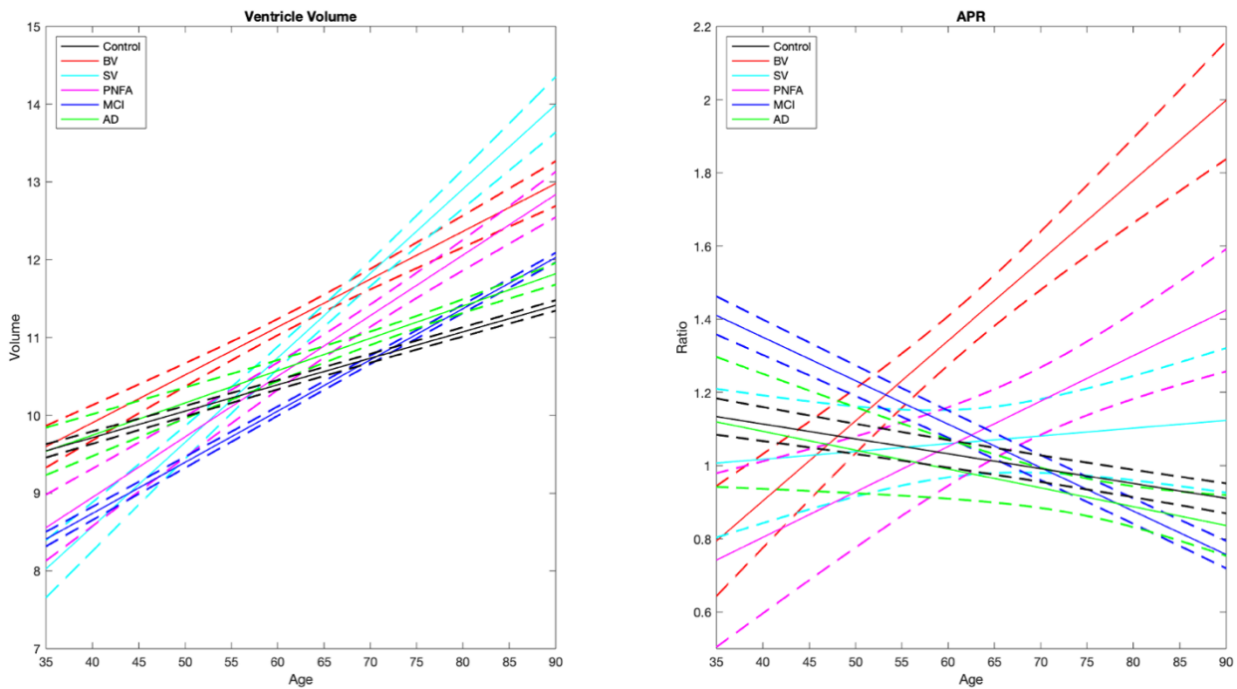

**Figure 4.** Plot showing the interaction between cohort and age for total ventricular volume and antero-posterior ratio. Model: Ratio Antero-Posterior ~ 1 + Cohort:Age + Sex + 1|ID . Abbreviations: MCI: mild cognitive impairment; AD: Alzheimer's disease; BV: behavioral variant; SV: semantic variant; PNFA; progressive non-fluent aphasia.

**Classification: bvFTD vs other dementias**

Using the ventricular APR to identify bvFTD from a mixed age-matched cohort (Control, MCI, AD, SV and PNFA) yielded a 10-fold cross-validation accuracy of 76±0.03 % (accuracy%±std) along with 72% sensitivity and 79% specificity. Adding an additional feature such as total VV resulted in an accuracy of 80±0.03 % (78% sensitivity and 82% specificity), while left/right temporal ratio or left/right frontal ratio did not improve global classification performances (77±0.03 % and 75±0.03 % respectively). Using all these features together (APR + TVV + LRTR + LRFR) improved the performances in bvFTD vs all cohorts classification (accuracy 80±0.03%, sensitivity 76% and specificity 83%).

The top accuracies against each individual cohort were 83±0.02 % (81% sensitivity, 87% specificity) for bvFTD vs Control; 89±0.02 % (87% sensitivity and 91% specificity) for bvFTD vs MCI using APR + Total VV and 83±0.01 % (81% sensitivity and 84 % specificity) for bvFTD vs AD using APR + LRTR. The best accuracy discriminating bvFTD from SV and PNFA were 66±0.03 % (60% sensitivity and 72% specificity) and 71±0.04 % (75% sensitivity and 68% specificity) respectively, using APR + TVV + LRTR + LRFR. (Figure 4).

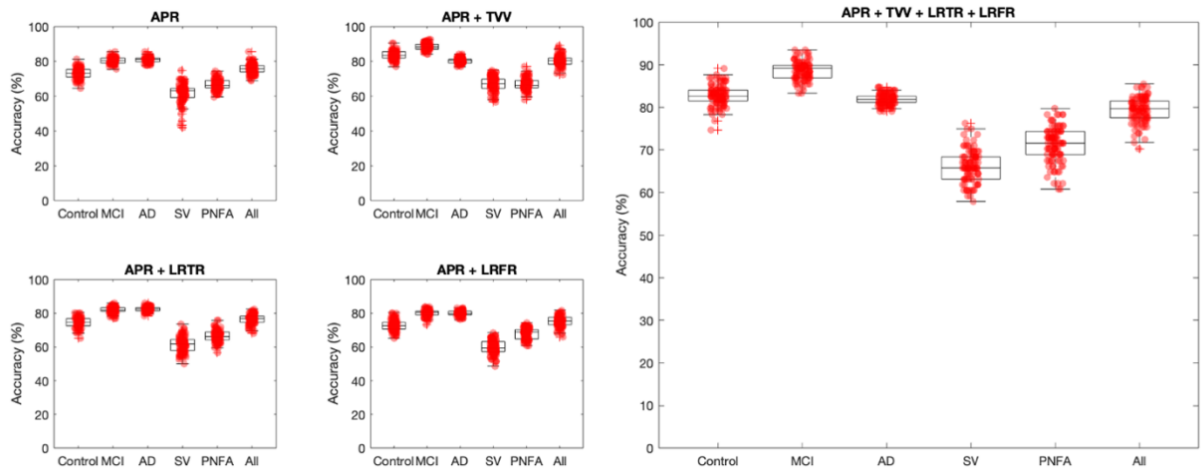

**Figure 5.** Boxplots showing the mean 10-fold classification accuracy, sensitivity and specificity with their 95% confidence intervals for bvFTD vs each individual cohort and the whole mixed dataset using *age + sex + anteroposterior ventricular ratio* alone and together with other volumetric ventricular features.

Abbreviations: MCI: mild cognitive impairment; AD: Alzheimer's disease; BV: behavioral variant; SV: semantic variant; PNFA; progressive non-fluent aphasia. APR: anteroposterior ratio; TVV: total ventricular volume; LRTR: left/right temporal ratio; left/right frontal ratio.

**bvFTD Classification Tool**

The ventricle feature estimation and classification tool developed in this project (VentRa) is publicly available at: http://nist.mni.mcgill.ca/?p=2498. VentRa takes a comma separated (.csv) file providing the path for the raw T1-weighted images as well as age and sex of the subjects as input, and provides preprocessed images along with ventricle segmentations, QC files for the segmentations, as well as a .csv file including the diagnosis (based on the classifier trained on bvFTD vs the mixed group data) along with all the extracted ventricle features: i.e. total ventricle volume, ventricle volumes in each lobe and hemisphere, APR, LRTR, and LRFR ratios. To provide some examples showing the performance of the classifier, average templates of control, bvFTD, SV, and PNFA [19] are also included in the package, with the generated outputs. VentRa requires MATLAB and minctools, the latter available at https://github.com/BIC-MNI/minc-toolkit-v2.

**DISCUSSION**

In this study, we investigated 1) the patterns of ventricular enlargement in bvFTD, SV, PNFA, MCI, and AD patients, and 2) the utility of ventricle-based features in differentiating bvFTD from CN, SV, PNFA, MCI, and AD. Our results showed a consistent pattern of ventricle enlargement in the bvFTD patients, particularly in the anterior parts of the frontal and temporal horns of the lateral ventricles. Temporal horns also exhibit the greatest enlargement in AD and SV compared to CN subjects.

Although some degree of ventricle expansion is expected with aging, the APR showed a much greater change with age than the total VV for the bvFTD group, in comparison with normal controls as well as with other dementia cohorts. Importantly, the significant increase in the APR is essentially a specific bvFTD feature since for other cohorts APR has minimal increase, remains stable or even decrease (i.e., other diseases impact posterior areas to a greater extent).

Ventricular APR was able to differentiate bvFTD from a mixed age-matched cohort (CN, MCI, AD, SV and PNFA) with an accuracy of 76% with high specificity. Furthermore, we achieved 89% and 83% accuracy distinguishing bvFTD from amyloid + MCI and AD respectively, together with 66-71% accuracy for bvFTD vs other FTD variants. Of note, the specificity and sensitivity were over 80 or 90% for all the non-FTD classifications (bvFTD vs controls, bvFTD vs amyloid + MCI and bvFTD vs amyloid + AD), which is the most clinically relevant. This performance is similar to the best performance reported in several articles that have analyzed structural MRI features [20-24]. Using an anteroposterior index derived from the relation between cortical atrophy within anterior and posterior regions, Bruun et al. reported areas under the curve (AUC) values of between 0.82 and 0.85 when separating all variants frontotemporal dementia (FTD) from AD in mixed group of non-FTD dementias (AD + Lewy body disease + vascular dementia + MCI + subjective cognitive decline + others)[22]. These results are in accordance with the accuracies in the present study, supporting the diagnostic value of anteroposterior atrophy gradient. However, we used ventricular volumes and ratios as proxy of brain atrophy, making it more sensitive for detection of subcortical atrophy. The lower accuracies obtained when identifying bvFTD from SV and PNFA could be related to the well-known clinical overlap amongst all entities in the FTLD spectrum, in particular bvFTD with language variants of FTD.

To ensure that minor inaccuracies in the ventricle segmentations did not impact our results, the ventricle segmentations were strictly QCed, and the results that did not pass this QC step were manually corrected for the rest of the analyses. However, to investigate how much such inaccuracies might impact the performance of the classifier, we also repeated the classification experiments with the uncorrected segmentations, and obtained similar results (i.e., 79% accuracy, 77% sensitivity and 81% specificity for bvFTD vs all other classification).

Our results suggest that APR might be a useful feature to aid bvFTD diagnosis, particularly given that the lateral ventricles can be reliably segmented using a variety of publicly available tools such as FreeSurfer and the patch-based method used in this paper[8, 9].

Our study has some limitations. The FTD patients were obtained from the FTLDNI dataset, whereas the MCI and AD patients were obtained from the ADNI dataset, raising the question of whether differences in images between the two studies might have impacted the findings. The likelihood of such difference is low, given that FTLDNI and ADNI use similar scanning protocols. Further, all of the image processing tools used in this study have been established and validated for use in multi-center and multi-scanner datasets [15, 25-29], and have been designed to minimize such differences. In addition, all the voxel-wise analyses (which were most likely to be affected by such differences) compared the patient groups versus normal control participants from the same study; i.e. FTLDNI bvFTD, SV, and PNFA patients were compared against FTLDNI CNs, and MCI and AD patients from ADNI were compared against ADNI CNs. The ventricle features are much less likely to be impacted by such differences, particularly for the APR, where any such differences would be cancelled out in the ratio.

The actual relevance of a biomarker aimed to distinguish behavioral vs language variants is limited for many of the cases where it is clinically evident. Yet, it could still be useful for the differential diagnosis in subjects that simultaneously fulfill clinical criteria for bvFTD and primary progressive aphasia (SV/PNFA) where brain imaging would be the hallmark. Similarly, while the value of a diagnostic algorithm to differentiate bvFTD from typical amnestic AD and MCI due to AD has limited clinical impact, such as tool has potential to facilitate the diagnosis of bvFTD in a clinical context against a broader differential diagnosis including primary psychiatric disorders. Given that primary psychiatric disorders are expected to show very modest volume lost at the most, the accuracy is expected to be close to the difference between bvFTD and CNs. This will have to be demonstrated in a mixed neuropsychiatric cohort in future work. It will also be interesting to determine the accuracy of the ratio between bvFTD and amyloid positive frontal-dsyexecutive AD. In general, the performances reported from visual radiologists' appear poorer than the classification accuracies achieved and they strongly rely on their level of experience[7], indicating the potential usefulness of an automated MRI-based tool for improving the diagnostic certainty of FTD. Moreover, the use of one single morphometric ratio would be simpler than multiple semi-structured visual rating scales of atrophy that are currently used in the clinic.

In conclusion, our study proposes an easy to obtain and generalizable ventricle-based feature (APR) from T1-weighted structural MRI (routinely acquired and available in the clinic) that can be used not only to differentiate bvFTD from normal subjects, but also from other FTD variants (SV and PNFA), MCI, and AD patients. In addition, we have made our ventricle feature estimation and bvFTD diagnosis tool (VentRa) publicly available, allowing application of our model in other studies. Of note, VentRa is not currently validated for clinical use. If validated in a prospective study, the proposed method has the potential to aid bvFTD diagnosis, particularly in settings where access to specialized FTD care is limited.

**Declaration of interests:**

Dr. Manera reports no disclosures

Dr. Dadar reports no disclosures

Dr. Collins is co-founder of True Positive Medical Devices.

Dr. Ducharme receives salary funding from the Fonds de Recherche du Québec - Santé. Dr. Ducharme is the co-founder of Arctic Fox AI.

**Acknowledgements**

MD is supported a scholarship from the Canadian Consortium on Neurodegeneration in Aging in which SD and RC are co-investigators as well as an Alzheimer Society Research Program (ASRP) postdoctoral award. The Consortium is supported by a grant from the Canadian

Institutes of Health Research with funding from several partners including the Alzheimer Society of Canada, Sanofi, and Women's Brain Health Initiative. This work was also supported by grants from the Canadian Institutes of Health Research (MOP-111169).

DLC receives funding from the Canadian Institutes of Health Research (MOP- 111169), les Fonds de Research Santé Quebec Pfizer Innovation fund, and an NSERC CREATE grant (4140438 - 2012). We would like to acknowledge funding from the Famille Louise & André Charron.

SD receives salary funding from the Fonds de Recherche du Québec – Santé (FRQS)

Data collection and sharing for this project was funded by the Alzheimer's Disease Neuroimaging Initiative (ADNI) (National Institutes of Health Grant U01 AG024904) and DOD ADNI (Department of Defense award number W81XWH-12-2-0012). ADNI is funded by the National Institute on Aging, the National Institute of Biomedical Imaging and Bioengineering, and through generous contributions from the following: AbbVie, Alzheimer's Association; Alzheimer's Drug Discovery Foundation; Araclon Biotech; BioClinica, Inc.; Biogen; Bristol-Myers Squibb Company; CereSpir, Inc.; Cogstate; Eisai Inc.; Elan Pharmaceuticals, Inc.; Eli Lilly and Company; EuroImmun; F. Hoffmann-La Roche Ltd and its affiliated company Genentech, Inc.; Fujirebio; GE Healthcare; IXICO Ltd.; Janssen Alzheimer Immunotherapy Research & Development, LLC.; Johnson & Johnson Pharmaceutical Research & Development LLC.; Lumosity; Lundbeck; Merck & Co., Inc.; Meso Scale Diagnostics, LLC.; NeuroRx Research; Neurotrack Technologies; Novartis Pharmaceuticals Corporation; Pfizer Inc.; Piramal Imaging; Servier; Takeda Pharmaceutical Company; and Transition Therapeutics. The Canadian Institutes of Health Research is providing funds to support ADNI clinical sites in Canada. Private sector contributions are facilitated by the Foundation for the National Institutes of Health24

(www.fnih.org). The grantee organization is the Northern California Institute for Research and Education, and the study is coordinated by the Alzheimer's Therapeutic Research Institute at the University of Southern California. ADNI data are disseminated by the Laboratory for Neuro Imaging at the University of Southern California.

Data collection and sharing for this project was funded by the Frontotemporal Lobar Degeneration Neuroimaging Initiative (National Institutes of Health Grant R01 AG032306). The study is coordinated through the University of California, San Francisco, Memory and Aging Center. FTLDNI data are disseminated by the Laboratory for Neuro Imaging at the University of Southern California.25